\documentclass[aps,pra,twocolumn]{revtex4-1}
\usepackage{graphicx}
\begin{document}
\title{Determining the parity of a permutation
using an experimental NMR qutrit}
\author{Shruti Dogra}
\email{shrutidogra@iisermohali.ac.in}
\address{Department of Physical Sciences, Indian
Institute of Science Education \&
Research (IISER) Mohali, Sector 81 SAS Nagar,
Manauli PO 140306 Punjab India.}
\author{Arvind}
\email{arvind@iisermohali.ac.in}
\address{Department of Physical Sciences, Indian
Institute of Science Education \&
Research (IISER) Mohali, Sector 81 SAS Nagar,
Manauli PO 140306 Punjab India.}
\author{Kavita Dorai}
\email{kavita@iisermohali.ac.in}
\address{Department of Physical Sciences, Indian
Institute of Science Education \&
Research (IISER) Mohali, Sector 81 SAS Nagar,
Manauli PO 140306 Punjab India.}
\begin{abstract}
We present the NMR implementation of a recently
proposed quantum algorithm  to find the parity of
a permutation.  In the usual qubit model of
quantum computation, speedup requires the presence
of entanglement and thus cannot be achieved by a
single qubit.  On the other hand, a qutrit is
qualitatively more quantum than a qubit because of
the existence of quantum contextuality and a
single qutrit can be used for computing.  We use
the deuterium nucleus oriented in a liquid crystal
as the experimental qutrit.  This is the first
experimental exploitation of a single qutrit to
carry out a computational task. 
\end{abstract}
\maketitle
\section{Introduction}
Quantum information processors exploit intrinsic
quantumness of quantum systems to perform
computational tasks more efficiently than their
classical 
counterparts~\cite{nielsen-book-02,ladd-nature-2010}.  
The notion of
quantumness of a quantum system has many aspects,
including quantum
entanglement~\cite{deutsch-prsla-1989,dorai-pra-2001},
contextuality~\cite{kirchmair-nature-2009,zu-prl-2012,amselem-prl-2012,
cabello-pra-2012},
nonlocality~\cite{lapkiewicz-nature-2011} and
fragility toward quantum
measurements~\cite{kurzynski-pra-2012}.  A
qubit is the building block of a standard quantum
computer, however a single qubit does not possess
intrinsic quantum features. In fact, the
polarization states of a classical beam of light
provide a classical system with exactly the same
properties as that of a single qubit. Therefore,
by manipulating the polarization states of a
classical beam of light via half-wave and
quarter-wave plates, one can efficiently simulate
a single
qubit~\cite{arvind-prjp-2001-2,arvind-josab-2007}.
In a quantum computer, we invariably have
multiple qubits and can create highly entangled states
of these qubits which are exploited to perform
computation~\cite{deutsch-prsla-1992,arvind-prjp-2001}.
There is no classical physical system on which we
can generate such entangled states because the
concept of multi-qubit entanglement is an intrinsically
quantum phenomenon. Therefore, although a single
qubit cannot be used to do any effective quantum
computation, multiple qubits can in fact provide a
computational resource.  Scaling issues in
algorithmic implementations of the and achieving
computational speedup without entanglement has
been discussed
in~\cite{dorai-currs-2000,arvind-pra-2003,
collins-pra-2010}.  On the contrary, a single
qutrit (a three-level quantum system) has no
classical counterpart~\cite{li-sr-2013}. It is the
simplest quantum system where the notion of
contextuality can be
introduced~\cite{thompson-sr-2013}. Quantum
contextuality is a strange property of quantum
systems where measurement outcomes of an
observable $A$ depend on the ``context'' in which
we perform the measurement, the context being the
set of other observables which are measured
along with $A$ (that commute with $A$ and among
themselves). For a qutrit, for a given observable,
we can have more than one context, and therefore a
qutrit possesses contextuality. In contrast, for a
qubit, for every observable there is only one
context. This makes a case for considering a
single qutrit as a possible resource for quantum
computation. A recent paper explores this
possibility in a concrete context of determining
the parity of a given permutation of three
objects~\cite{gedik-qph-2014}. They have shown in
their model algorithm that for the six possible
permutations (three even and three odd),
the parity can be determined by a single call to
the quantum oracle, as opposed to two calls in the
classical setup. The algorithm can be
generalized to higher-dimensional qudits with
the same two to one speedup ratio.  This 
work is an interesting
development because it uses 
the quantumness of a single
system, where there is no
question of entanglement, to perform a
computation.
The author also alludes
to the possibility of the contextuality of a qutrit
playing a role in this speedup.  Recent work on
quantum information processing with qutrits
includes the description of unitary
gates~\cite{levaillant-qph-2013}, a quantum
circuit to solve systems of linear
equations~\cite{sawerwain-qph-2013}, a fundamental
test of
nonclassicality~\cite{lapkiewicz-qph-2013,ahrens-sr-2013},
the measurement of entangled
qutrits~\cite{langford-prl-2004}, experimental
quantum cryptography~\cite{groblacher-newj-2006},
to study entanglement sudden
death~\cite{ann-pla-2008},
and an experimental test of quantum
contextuality~\cite{zhang-prl-2013}.

In this work we implement the parity determining
quantum algorithm on an NMR quantum information
processor.  Recent work on qudit computation using
NMR includes a study of relaxation dynamics in a
quadrupolar system~\cite{auccaise-jmr-2008},
the preparation of pseudopure states in a single
qutrit~\cite{das-ijqi-2003}, the 
emulation of a qutrit-qubit system using
strongly dipolar coupled spin-1/2 
particles~\cite{gopinath-pra-2006},
and state tomography and logical
operations in a three qubits NMR
quadrupolar system~\cite{ferreira-ijqi-2012}.  
Here we
have used a deuterium
(spin-1 particle) nucleus  as the
NMR qutrit.
When a
strong magnetic field is applied, the three levels
of the spin-1 deuterium nucleus Zeeman split and
provide us with two degenerate transitions. The
degeneracy is lifted by orienting the system in a
liquid crystal matrix where the anisotropic
environment leads to a contribution of the dipolar
coupling to the Hamiltonian.  To the best of our
knowledge, this is the first experimental
exploitation of the quantumness of a single qutrit
for quantum computation.

The material in this paper is arranged as follows:
Section~\ref{algorithm} describes the parity
determination quantum algorithm given in
~\cite{gedik-qph-2014}. Section~\ref{nmr} contains
a description of the  quantum circuit constructed
to experimentally implement the parity algorithm,
details of the experimental NMR qutrit, and a
discussion of the results.  Section~\ref{concl}
contains some concluding remarks.

\section{Parity Determining Algorithm}
\label{algorithm}
A qutrit-based black-box algorithm has been
recently designed to evaluate the parity of
permutation of three objects by one oracle call,
whilst the classical algorithm requires two oracle
calls~\cite{gedik-qph-2014}. Since the algorithm
uses a single qutrit and has a quantum advantage,
the authors conjecture that this computational
speedup may be attributed to contextuality, an
intrinsic  quantum feature present in a qutrit.

Consider the six possible permutations of the three objects
in the set $\{1,2,3\}$, categorized as {\bf even} or {\bf
odd}, depending on whether the number of exchange operations
performed is {\bf even} or {\bf odd}.  The computational
task that one wants to perform is to find the parity of the
permutation. If we consider the permutation as a function $f(x)$
where $x\in \{1,2,3\}$, then to determine the parity
classically we need to evaluate $f(x)$ for at least two
values of $x$.

In the quantum setting, let us consider a qutrit with its 
eigen states labeled by  its spin quantum number
$\vert m \rangle$ where $m=1,0,-1$. The action
of permutations on these states are the bijective maps
$f : \{1,0,-1\} \longrightarrow \{1,0,-1\}$. The six possible
maps can be written down using Cauchy's notation. The three
even maps and the corresponding unitary transformations are 
given by:
\begin{eqnarray}
f_1 = \left(\!\!\!\! \begin{array}{rrr}
                  \phantom{-}1 & \phantom{-}0 & -1 \\
                  1 & 0 & -1
                 \end{array}  \right);
&& 
U_1=\left(
\begin{array}{ccc}
1 & 0 & 0\\
0 & 1 & 0 \\
0 & 0 & 1
\end{array}
\right)
\nonumber \\
f_2 = \left(\!\!\!\! \begin{array}{rrr}
                  \phantom{-}1 & 0 & -1 \\
                  0 & -1 & 1
                 \end{array} \right) 
; &&
 U_2=\left(
\begin{array}{ccc}
0 & 1 & 0\\
0 & 0 & 1 \\
1 & 0 & 0
\end{array}
\right)
\nonumber \\
f_3 = \left(\!\!\!\! \begin{array}{rrr}
                  1 & \phantom{-}0 & -1 \\
                  -1 & 1 & 0 
                 \end{array} \right)
; &&
 U_3=\left(
\begin{array}{ccc}
0 & 0 & 1\\
1 & 0 & 0 \\
0 & 1 & 0
\end{array}
\right)
\end{eqnarray}

while the three odd maps and the corresponding unitary
transformations are given by:
\begin{eqnarray}
f_4 = \left(\!\!\!\! \begin{array}{rrr}
                  \phantom{-}1 & \phantom{-}0 & -1 \\
                  0 & 1 & -1
                 \end{array} \right) 
; &&
 U_4=\left(
\begin{array}{ccc}
0 & 1 & 0\\
1 & 0 & 0 \\
0 & 0 & 1
\end{array}
\right)
\nonumber \\
f_5 = \left(\!\!\!\! \begin{array}{rrr}
                  \phantom{-}1 & 0 & -1 \\
                  1 & -1 & 0 
                 \end{array} \right)
; &&
 U_5=\left(
\begin{array}{ccc}
1 & 0 & 0\\
0 & 0 & 1 \\
0 & 1 & 0
\end{array}
\right)
\nonumber \\
f_6 = \left(\!\!\!\! \begin{array}{rrr}
                  1 & \phantom{-}0 & -1 \\
                  -1 & 0 & 1
                 \end{array}  \right)  
; &&
 U_6=\left(
\begin{array}{ccc}
0 & 0 & 1\\
0 & 1 & 0 \\
1 & 0 & 0
\end{array}
\right)
\end{eqnarray}

The quantum oracle thus is a black box which performs 
the unitary transformation corresponding to a given
permutation on the input state of the qutrit.

A direct run of the oracle on the eigen states of the qutrit
obviously does not help, and we would still
require two oracle calls,
as in the classical case.  The quantum algorithm
thus begins by
preparing a superposition of eigen states by the action of
the quantum Fourier transformation for a qutrit
acting on an initial
state (we consider 
the state $\vert -1\rangle$) and defined through the unitary
operation
\begin{equation}
F = \frac{1}{\sqrt{3}}\left(
\begin{array}{ccc} 1 & 1 & 1 \\ 1 &
e^{\frac{2\pi\iota}{3}} & e^{\frac{-2\pi\iota}{3}}
\\ 1 & e^{\frac{-2\pi\iota}{3}} &
e^{\frac{2\pi\iota}{3}} \end{array}  \right)
\label{FT}
\end{equation}
This is followed by the oracle call which is a black box
operation where the  unitary operator $U_f$ (corresponding
to the function $f_k$) is applied to the superposition state
created by the quantum Fourier transformation.  The output
state is then subjected to an inverse Fourier transformation
which brings the state back to one of the  eigen states of
the qutrit up to a phase factor.  The final state of the
qutrit is either $\vert - 1 \rangle$ (if $f_k$ is even) or
$\vert 0 \rangle$ (if $f_k$ is odd).  The final state for
the case of even functions is orthogonal to the final state
for the case of odd function and are thus distinguishable by
a measurement on a single copy. The result of such a
measurement in the $\{\vert 1\rangle, \vert 0\rangle,
\vert -1 \rangle\}$ basis  will thus reveal the parity of
the corresponding permutation.
\begin{figure}[ht]
\centerline{
\includegraphics[angle=0,scale=1.0]{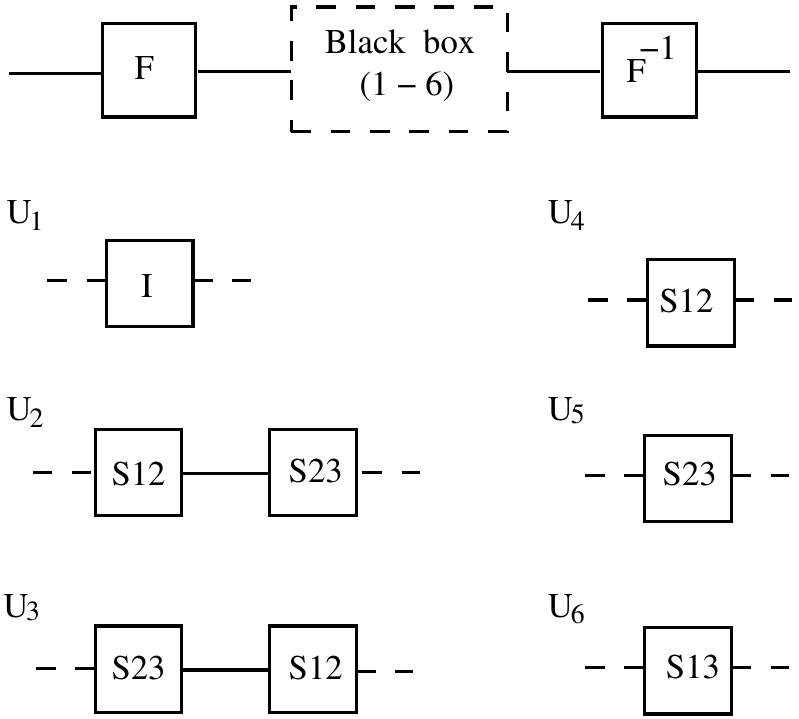}}
\caption{
Quantum circuit to determine the parity of the permutation
of three objects in a single step.  The initial state is the
pure state of a single qutrit, $\vert -1 \rangle$. Fourier
transformation and its inverse are represented as '$F$' and
'$F^{-1}$'.  The black box carrying out the permutations has
six different possibilities. '$I$' is the identity operator,
while '$S12$', '$S23$' and '$S13$' are SWAP operators
describing a swap between levels $1-2$, $2-3$ and $1-3$ of a
single qutrit respectively.}
\label{circuit}
\end{figure}

The unitaries corresponding to different functions can be
constructed using combinations of SWAP operators as shown in
Figure~\ref{circuit}. The  even unitaries are
constructed by the identity operator (no
operation), and by two sequential non-commuting swap
gates $S_{12}-S_{23}$ and $S_{23}-S_{12}$, 
where the $S_{ij}$ gate performs a swap
between the $i$th and $j$th levels. The odd
unitaries are constructed by the single swap gates
$S_{12}, S_{23}$ and $S_{13}$ respectively.
\section{NMR Implementation of the Parity Algorithm}
\label{nmr}
\subsection{The NMR Qutrit}
\begin{figure}[ht]
\centerline{
\includegraphics[angle=0,scale=1.0]{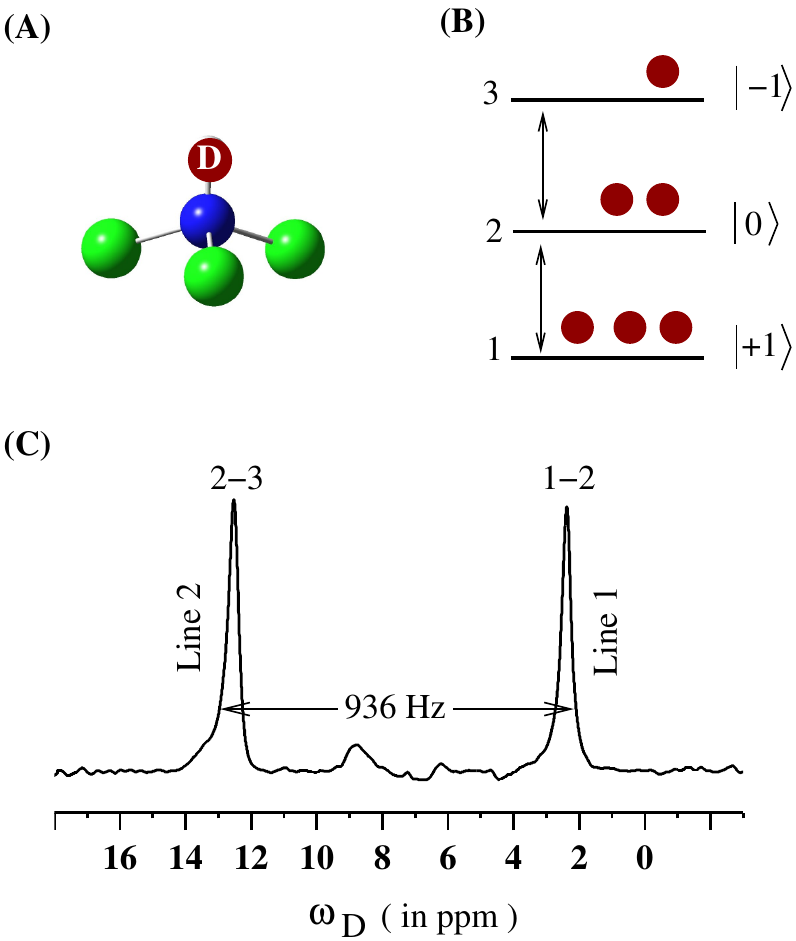}}
\caption{
A single NMR qutrit is constructed by orienting
deuterated Chloroform in a lyotropic liquid
crystal, with the deuterium, (spin 1) being the
single qutrit: (a) the Chloroform-D molecule, (b)
the energy level diagram of a single qutrit.
Energy levels are numbered as 1,2,3 and the
corresponding basis vectors are represented on the
right. Relative populations of the energy levels
at equilibrium are shown with red circles.  Two
spectral lines in an NMR spectrum of single qutrit
result from the transitions between energy levels
1-2 and 2-3 as shown with arrows. (c) The
deuterium NMR spectrum of Chloroform-D  with the
transitions labeled as Line 1 and Line 2.  The
experiments were performed at $277$ K resulting in
a quadrupolar splitting of 936 Hz.
\label{system}}
\end{figure}
We use the deuterated chloroform molecule oriented
in a lyotropic liquid crystal as our system, where
the deuterium nucleus serves as the single qutrit.
The lyotropic liquid crystal is composed of $25.6
\%$ of Potassium Laurate, $68.16 \%$ of $H_2O$ and
$6.24 \%$ of Decanol, and 50 $\mu l$ of
Chloroform-D is added to 500 $\mu l$ of liquid
crystal.  The deuterium NMR spectrum of oriented
Chloroform-D was acquired at different
temperatures, as this system possesses a liquid
crystalline phase for a wide range of temperature.
Figure~\ref{system} shows the
energy level diagram of the three level system
along with the relative populations in the presence
of a strong magnetic field $B_0$. The energy
levels are numbered as $\{1,2,3\}$ 
corresponding to the qutrit eigenstates
$\vert +1 \rangle, \vert 0 \rangle, \vert -1
\rangle$ respectively,
and the single
quantum transitions are labeled with arrows.
Figure~\ref{system}(c) shows the deuterium NMR spectrum of
oriented Chloroform-D; the spectral lines
corresponding to transitions 1-2 and 2-3 are
labeled as Line 1 and Line 2 respectively. 

For a deuterium nucleus with spin $I = 1$, the electric
quadrupole moment of the nucleus interacts with
the electric field gradients generated by the
surrounding electron cloud, leading to a
quadrupolar coupling term in the 
Hamiltonian~\cite{levitt-book-2008,slichter-book-1996} 
\begin{equation}
H = - \omega_0 I_z + \frac{e Q
V_{zz}}{4I(2I-1)}(3I_z^2-I^2)) 
\end{equation}
where the first term describes the Zeeman
interaction and 
the second term describes the first-order 
quadrupolar interaction,
$Q$
is the quadrupolar moment and $V_{zz}$ is the
average value of the uniaxial
electric field gradient component over the
molecular motion.

In an isotropic environment the electric
field gradient component $V_{zz}$ averages to zero, 
and the quadrupolar term in the Hamiltonian
vanishes. 
The two
single-quantum transitions become degenerate and
thus cannot be distinguished. 
For quantum computing and gate implementation on 
the deuterium qutrit, one requires
the two single-quantum transitions to be separately
manipulatable, which is not possible 
with liquid-state NMR methods.
To resolve this problem, the molecule
is embedded
in a liquid crystalline environment and the
anisotropic molecular orientation 
with respect to the strong magnetic field, gives rise to a finite
quadrupolar coupling term in the Hamiltonian    
now given by
\begin{equation}
H = - \omega_0 I_z
+ \Lambda (3 I_z^2 - I^2) 
\end{equation}
where 
$\Lambda = e^2 q Q S / 4$ is 
the effective value of the quadrupolar coupling,
$e q =V_{zz}$ is denoted the field gradient parameter,
and $S$ is the order parameter of the liquid
crystal.
This effective quadrupolar coupling term is
responsible for the splitting between the
previously degenerate energy levels and
hence for
the two non-degenerate lines in the 
NMR spectrum. This term could be very large
(in MHz) but its contribution is tunable by the
order parameter $S$ of the liquid crystal, and
a small value of $S$ reduces the effective
quadrupolar coupling to a few hundred Hz.
At 277 K, the effective quadrupolar splitting
between the spectral lines was measured to be $936
Hz$, and all the experiments were 
performed at this temperature.

\subsection{Experimental Results}
\begin{figure}[ht]
\centerline{
\includegraphics[angle=0,scale=1]{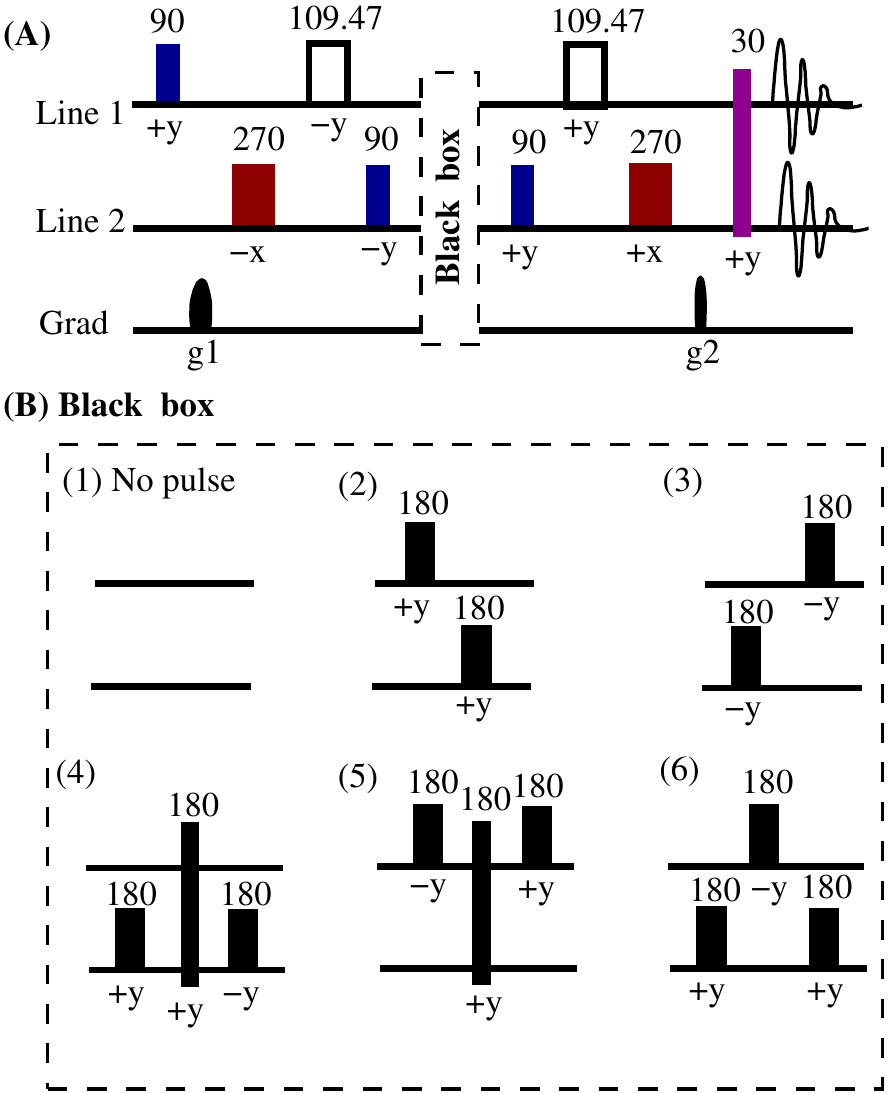}}
\caption{
(a) Pulse sequence for the implementation of the
parity algorithm on a single qutrit.  The first
two channels of the pulse sequence correspond to
the two NMR transitions.  The first $90^0$ pulse
on Line 1 followed by a gradient g1 creates the
pseudopure state $\vert -1 \rangle$. The next
three pulses perform the Fourier transformation.
(b) The black box carries out the six possible
permutations corresponding to the three even and
three odd functions.  The two channels inside the
black box correspond to the two NMR transitions of
a single qutrit.  A $30^0$ non-selective detection
pulse preceded by a clean up gradient g2 is
applied to evaluate the final result of the
computation.  All the $90^0$ pulses are shown in
blue, $270^0$ in red and $180^0$ in black.
Non-selective pulses are shown as a common
rectangle for both the transitions. Pulse angles
and the axes of rotations are shown corresponding
to each pulse.  All the pulses are shaped pulses.
\label{pulse}}
\end{figure}
All the experiments were performed at 277 K on a 600 MHz Avance
III NMR spectrometer equipped with a QXI probe,
with the deuterium nucleus resonating at 91.108 MHz.  
The deuterium relaxation times $T_1$ and $T_2$
were of the order of 170 ms and 50 ms
respectively. 
The system
is initialized into the pseudopure state $\vert -1 \rangle$
which is obtained by applying a $90^0$ pulse on transition
1-2 which equalizes the populations
of levels 1 and 2, followed by a z-gradient pulse (g1 in
Figure~\ref{pulse}). 
A Fourier transformation is then implemented by a sequence of three
transition-selective pulses 
$(270)_{-x23} \, (109.47)_{-y12} \, (90)_{-y23}$.
This leads to the superposition
state
$\frac{1}{\sqrt{3}}(\vert +1 \rangle+e^{\frac{-2\pi
\iota}{3}}\vert 0 \rangle+e^{\frac{2\pi
\iota}{3}}\vert -1\rangle)$. 
When this state undergoes odd permutations
we obtain the
states $\frac{1}{\sqrt{3}}(e^{- 2 \pi \iota/3}
\vert +1 \rangle +  \vert 0 \rangle + e^{2
\pi \iota/3} \vert -1 \rangle)$,
$\frac{1}{\sqrt{3}}(\vert +1 \rangle 
+ e^{2 \pi \iota/3}
\vert 0 \rangle + e^{-2 \pi \iota/3}  \vert -1
\rangle)$ and
$\frac{1}{\sqrt{3}}(e^{2 \pi
\iota/3} \vert +1 \rangle + e^{-2 \pi \iota/3}
\vert 0 \rangle +
\vert -1 \rangle)$ respectively.
Inverse Fourier transform of these states result
in the states $e^{-2 \pi
\iota/3} \vert 0 \rangle$, $\vert 0 \rangle$ and $e^{2 \pi
\iota/3} \vert 0 \rangle$ respectively.
For even permutations, the initial state is
unchanged, upto a global phase.
The permutations are implemented by various combinations of
$180^0$ transition-selective as well as non-selective
pulses as detailed in Figure~\ref{pulse}(b). 
The non-selective pulses act
on both the transitions and are shown as a common rectangle
on both the lines.
The final state is detected by a $30^0_y$
non-selective detection pulse preceded by a clean up
gradient g2.  
All the transition-selective pulses used in
this work are $4$ ms long 'Gaussian' shaped pulses and the
non-selective ones are 'Sinc' shaped pulses with a duration
of $0.5$ ms. There is no gradient implementation while
permuting the elements. The entire set of pulses 
used to perform the algorithm were
implemented before decoherence sets in.
\begin{figure}[ht]
\centerline{\includegraphics[angle=0,scale=1.0]{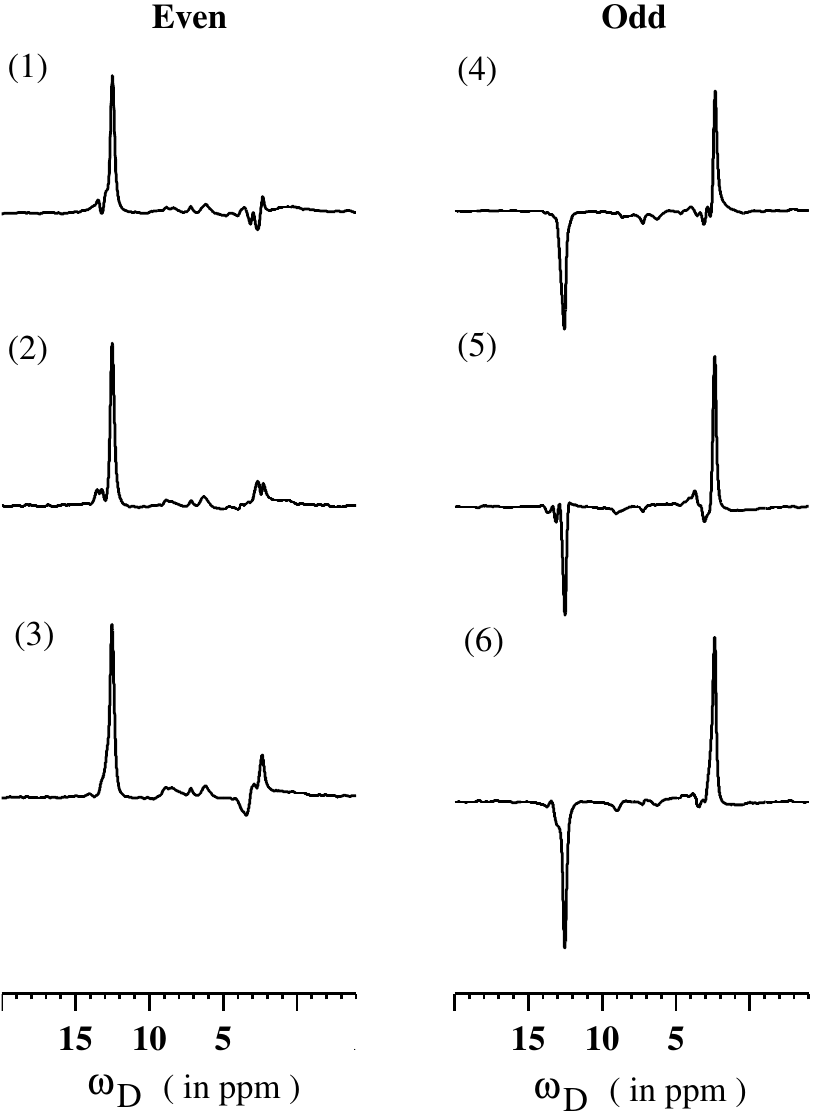}}
\caption{
The NMR spectra after implementing
the parity determining algorithm on a
single qutrit.
Spectra (numbered 1 to 6)  correspond to the six
possible permutations.  All the spectra were
obtained by applying a $30^0$ non-selective
detection pulse on both the transitions. Spectra
1,2,3 correspond to the state $\vert -1 \rangle$
(resulting from an even permutation) and spectra
4,5,6 correspond to state $\vert 0 \rangle$
(resulting
from an odd permutation).
\label{spectrum}}
\end{figure}

The deuterium NMR spectrum obtained after the six
different permutations is shown in
Figure~\ref{spectrum}.  There are two possible
resultant states after the computation is
performed, depending upon whether the permutation
is even or odd.  If the permutation is even, the
final state is $e^{\iota \phi} \vert -1 \rangle$
and if it is odd, the final state is $e^{\iota
\phi} \vert 0 \rangle$.  These two final states
are orthogonal to each other and can hence be
distinguished by a single projective measurement.
Since projective measurements are not possible on
an ensemble NMR quantum computer, we use their
spectral signature to distinguish these two
orthogonal states.  
The state $\vert -1 \rangle$,
when detected by a  
applying a $30^0$ non-selective
pulse on both the transitions, gives rise to a line
corresponding to transition 2-3 and the state $\vert 0
\rangle$ with the same detection pulse gives two
lines at transitions 1-2 and 2-3 with equal and
opposite intensities.  Global phases are not
detectable in NMR experiments and can hence be
ignored.

Figure~\ref{spectrum} depicts
the phase corrected spectra (at the same
intensity scale) corresponding to six different
permutations. The  
spectra on the left correspond to the final state
$\vert -1 \rangle$ after an even
permutation, while the spectra on the
right correspond to
the final state
$\vert 0 \rangle$ after an odd permutation. 

\section{Concluding Remarks}
\label{concl}
We have experimentally implemented a parity
determining algorithm on a single NMR qutrit,
which does not require entanglement to achieve
computational speedup.  We have oriented the
deuterium qutrit in a liquid crystalline
environment bringing in a difference in the
resonance frequencies of the two transitions,
which was a key component in allowing us to
manipulate the qutrit states.  Although the
demonstration of the computing power of a single
qutrit has been shown using only a toy algorithm,
this experiment paves the way for future work in
this area, whereby the contextuality of qutrits can
be harnessed to implement useful quantum
algorithms, in much the same way as entanglement
is used as a computational resource in multi-qubit
systems.  Interestingly, the dynamics of a single
qubit can always be efficiently simulated
classically, and
therefore does not lead to a computational
advantage for any quantum algorithm. A single
qutrit on the other hand, possesses contextuality,
and can be used for quantum computation.

\vspace*{1cm}
\noindent{\bf Acknowledgments}
All experiments were performed on the Bruker 600 MHz
FT-NMR spectrometer in the NMR Research Facility
at IISER Mohali. SD was funded by a Government of
India UGC-SRF fellowship.
\vspace*{1cm}

\end{document}